\documentclass[preprint,authoryear,12pt]{elsarticle}
\usepackage[round]{natbib}
\usepackage{amssymb,amsthm,amsmath}
\journal{ArXiv}

\begin{document}
\newtheorem{thm}{Theorem}
\newtheorem{cor}{Corollary}
\newproof{pf}{Proof}

\begin{frontmatter}

\title{Two stage design for estimating the product of means with cost in
the case of the exponential family}
\author[1]{Zohra BENKAMRA}
\author[2]{Mekki TERBECHE}
\author[1]{Mounir TLEMCANI\corref{*}}

\address[1]{Department of Mathematics. University of Oran, Algeria}%
\address[2]{Department of Physics, L.A.A.R University Mohamed Boudiaf, Oran, Algeria}
\cortext[*]{Corresponding author : mounir.tlemcani@univ-pau.fr (M.Tlemcani)}

\begin{abstract}
We investigate the problem of estimating the product of means of independent
populations from the one parameter exponential family in a Bayesian
framework. We give a random design which allocates $m_{i}$ the number of
observations from population $P_{i}$ such that the Bayes risk associated
with squared error loss and cost per unit observation is as small as
possible. The design is shown to be asymptotically optimal.
\end{abstract}


\end{frontmatter}%

\section{Introduction}

Assume that for $i=1,...,n$; a random variable $X_{i}$ whose distribution
belongs to the one parameter exponential family is observable from
population $P_{i}$ with cost $c_{i}$ per unit observation. The problem of estimating
several means in the case of the exponential family distributions with linear combination of losses
was addressed by \cite{cohen}.
The problem of interest in this paper is to estimate the product of means using a Bayesian approach
associated with squared error loss and cost. Since a Bayesian framework is
considered; see, e.g., \cite{page,shapiro}, then typically optimal estimators are
Bayesian estimators and the problem turns to design a sequential allocation
scheme; see, e.g., \cite{woodroofe}, to select $m_{i}$ the number of observations
from population $P_{i}$ such that the Bayes risk plus the corresponding
budget $B=\sum_{i=1}^{n}c_{i}m_{i}$ is as small as possible. \cite{terbeche aas},
have defined a sequential design to estimate the difference
between means of two populations from the exponential family with associated
cost. The random allocation was shown to be the best from numerical
considerations; see, e.g., \cite{terbeche phd}.
Similarly, the problem of estimating the product of several means of
independent populations, subject to the constraint of a total number of
observations $M$ fixed, was addressed by \cite{rekab}, using a two stage
approach. The allocation of $m_{i}$ was nonrandom and the first order
optimality was shown for large $M$.

Suppose that $X_{i}$ has the distribution of the form%
\[
f_{\theta _{i}}(x_{i})\varpropto e^{\theta _{i}x_{i}-\psi (\theta _{i})},~x_{i}\in 
\mathbb{R}
,~\theta _{i}\in \Omega
\]%
where $\Omega $ is a bounded open interval in $\mathbb{R}$.
It follows that $E_{\theta _{i}}\left[ x_{i}\right] =\psi ^{\prime }\left( \theta
_{i}\right) $ and $Var_{\theta _{i}}\left[ x_{i}\right] =\psi ^{\prime
\prime }\left( \theta _{i}\right)$. One assumes that prior distribution
for each $\theta _{i}$ is given by%
\[
\pi _{i}\left( \theta _{i}\right)\varpropto e^{r_{i}\left( \mu _{i}\theta
_{i}-\psi \left( \theta _{i}\right)\right) }
\]%
where $r_{i}$ and $\mu _{i}$ are reals and $r_{i}>0$, $i=1,...,n$. Here we
treat $\theta _{i}$ as a realization of a random variable and assume that
for each population, $x_{i1},...,x_{im_{i}}$ are conditionally independent
and that $\theta _{1},...,\theta _{n}$ are a priori independent. Our aim is to estimate the product
$\theta =\prod_{i=1}^{n}\psi ^{\prime }\left( \theta _{i}\right)$, subject to squared error loss and linear cost.

\section{The Bayes risk}

Let $\mathcal{F}_{m_{1},...,m_{n}}$ the $\sigma $-Field generated by $\left(
X_{1},...,X_{n}\right) $ where $X_{i}=\left( x_{i1},...,x_{im_{i}}\right) $
and let $\mathcal{F}_{m_{i}}=\sigma \left( X_{i}\right) =\sigma \left(
x_{i1},...,x_{im_{i}}\right) $.
It was shown that
\begin{eqnarray}
E\left[ \psi ^{\prime }\left( \theta _{i}\right) /\mathcal{F}_{m_{i}}%
\right] &=&=\frac{\mu_{i}r_{i}+\sum_{j=1}^{m_{i}}x_{ij}}{m_{i}+r_{i}}  \label{th 2.2.1-1} \\
Var\left[ \psi ^{\prime }\left( \theta _{i}\right) /\mathcal{F}_{m_{i}}%
\right] &=&E\left[ \frac{\psi ^{\prime \prime }\left( \theta _{i}\right) }{%
m_{i}+r_{i}}/\mathcal{F}_{m_{i}}\right],  \label{th 2.2.1-2}
\end{eqnarray}
 (see \cite{terbeche sort}). Using independence across populations, the Bayes estimator of $\theta $ is%
\[
\hat{\theta}=E\left[ \theta /\mathcal{F}_{m_{1},...,m_{n}}\right]
=\prod\limits_{i=1}^{n}E\left[ \psi ^{\prime }\left( \theta _{i}\right) /%
\mathcal{F}_{m_{i}}\right]
\]%
Assume that there exists $p\geq 1$ such that 
\begin{equation}
E\left[ \left( \psi ^{\prime \prime }\left( \theta _{i}\right) \right) ^{p}%
\right] <+\infty ~ and ~ E\left[ \left( \psi ^{\prime }\left( \theta
_{i}\right) \right) ^{2p}\right] <+\infty,  \label{cond}
\end{equation}%

for all $i=1,...,n$; then the corresponding Bayes risk associated with quadratic loss
and cost can be written as follows,%
\begin{equation}
\label{r}
R\left( m_{1},...,m_{n}\right) =E\left[ \sum\limits_{i=1}^{n}\frac{U_{im_{i}}%
}{m_{i}+r_{i}}+\sum\limits_{i=1}^{n}c_{i}m_{i}\right] +\sum%
\limits_{i=1}^{n}o\left( \frac{1}{m_{i}}\right)
\end{equation}
and by the way, it can be approximated for large samples by%
\begin{equation}
\label{rtilde}
\tilde{R}\left( P\right) =\tilde{R}\left( m_{1},...,m_{n}\right) =E\left[
\sum\limits_{i=1}^{n}\frac{U_{im_{i}}}{m_{i}+r_{i}}+\sum%
\limits_{i=1}^{n}c_{i}m_{i}\right]
\end{equation}
where $U_{im_{i}}=E\left[ V_{i}/\mathcal{F}_{m_{1},...,m_{n}}\right] $ and %
$V_{i}=\psi ^{\prime \prime }\left( \theta _{i}\right) \prod_{j\neq
i}\psi ^{\prime 2}\left( \theta _{j}\right)$

\section{Lower bound for the scaled Bayes risk}
From now on, the notation $c\rightarrow 0$ means that $c_{j}\rightarrow 0$, for all $j=1,...,n$.
Assume that for all $i$, 
\begin{equation}
\label{ci}
\frac{c_{i}}{\sum\limits_{j=1}^{n}c_{j}}\rightarrow \lambda _{i}\in \left]
0,1\right[ ,\ as \  c\rightarrow 0. 
\end{equation}
\begin{thm}
\label{th 233}For any random design (P) satisfying
\begin{equation}
\label{ai}
m_{i}\sqrt{c_{i}}\rightarrow a_{i}\neq 0~,~a.s., \ as\ c\rightarrow 0;
\end{equation}
then%
\begin{equation}
\label{res}
\liminf_{c\rightarrow 0}\frac{R(P)}{\sqrt{%
\sum\limits_{j=1}^{n}c_{j}}}\geq 2E%
\left[ \sum\limits_{i=1}^{n}\sqrt{\lambda _{i}}\sqrt{V_{i}}\right]
\end{equation}
\end{thm}

\begin{pf}
Expressions (\ref{r}) and (\ref{rtilde}) with the help of (\ref{ci}) and (\ref{ai}), give%
\begin{equation}
\label{rrtilde}
\liminf_{c\rightarrow 0 }\frac{R(P)}{\sqrt{%
\sum\limits_{j=1}^{n}c_{j}}}=\liminf_{c\rightarrow 0}\frac{\tilde{R}(P)}{\sqrt{\sum\limits_{j=1}^{n}c_{j}}}%
\end{equation}
and the scaled approximated Bayes risk satisfies the following inequality:%
\[
\frac{\tilde{R}(P)}{\sqrt{\sum\limits_{j=1}^{n}c_{j}}}\geq 2E\left[
\sum\limits_{i=1}^{n}\sqrt{\frac{c_{i}}{\sum\limits_{j=1}^{n}c_{j}}}\sqrt{%
U_{im_{i}}}\right] -\sum\limits_{i=1}^{n}\sqrt{\frac{c_{i}}{%
\sum\limits_{j=1}^{n}c_{j}}}\sqrt{c_{i}}r_{i},
\]
since for all $i$,
\begin{eqnarray*}
\frac{U_{im_{i}}}{m_{i}+r_{i}}+c_{i}m_{i} &=&\left( \frac{\sqrt{U_{im_{i}}}}{\sqrt{m_{i}+r_{i}}}+\sqrt{c_{i}}\sqrt{%
m_{i}+r_{i}}\right) ^{2}+2\sqrt{c_{i}}\sqrt{U_{im_{i}}}-c_{i}r_{i}\\
&\geq &2\sqrt{c_{i}}\sqrt{U_{im_{i}}}-c_{i}r_{i}.
\end{eqnarray*}

Finally, Fatou's lemma and condition (\ref{ci}) give%
\[
\liminf_{c\rightarrow 0}\frac{\tilde{R}(P)}{%
\sqrt{\sum\limits_{j=1}^{n}c_{j}}}\geq 2E\left[ \liminf_{c\rightarrow 0}\sum\limits_{i=1}^{n}\sqrt{\frac{c_{i}}{%
\sum\limits_{j=1}^{n}c_{j}}}\sqrt{U_{im_{i}}}\right] =2E\left[
\sum\limits_{i=1}^{n}\sqrt{\lambda _{i}}\sqrt{V_{i}}\right]
\]
and the proof follows.
\end{pf}

\section{First order optimal design}
According to condition (\ref{ai}) and identity (\ref{rrtilde}) a first order optimal design with respect to $\sum_{i=1}^{n}m_{i}=m$ must satisfy%
\begin{equation}
\label{conver}
\frac{\tilde{R}(P)}{\sqrt{\sum\limits_{j=1}^{n}c_{j}}}-2E\left[
\sum\limits_{i=1}^{n}\sqrt{\lambda _{i}}\sqrt{V_{i}}\right] \rightarrow 0, \ as\ c\rightarrow 0,
\end{equation}
It should be pointed that condition (\ref{conver}) is actually similar to the first order efficiency property for A.P.O rules in Bayes sequential
estimation; see, e.g., \cite{leng,hwang}, for one-parameter exponential families, which involves a sequential allocation procedure and a stopping time.

In our approach, condition (\ref{conver}) is handled by the following expansion.%
\begin{eqnarray*}
\frac{\tilde{R}(P)}{\sqrt{\sum\limits_{j=1}^{n}c_{j}}} &=& \frac{2E\left[ \sum\limits_{i=1}^{n}\sqrt{c_{i}}\sqrt{U_{im_{i}}}\right] }{%
\sqrt{\sum\limits_{j=1}^{n}c_{j}}}
+\frac{E\left[ \sum\limits_{i=1}^{n}\frac{\left( \sqrt{U_{im_{i}}}-\left(
m_{i}+r_{i}\right) \sqrt{c_{i}}\right) ^{2}}{m_{i}+r_{i}}\right] }{\sqrt{%
\sum\limits_{j=1}^{n}c_{j}}} \\
&-&\sum\limits_{i=1}^{n}\sqrt{c_{i}}\sqrt{\frac{%
c_{i}}{\sum\limits_{j=1}^{n}c_{j}}}r_{i}
\end{eqnarray*}%
The last term goes to zero as $c\rightarrow 0$, thanks to condition (\ref{ci}). Hence, sufficient conditions for a design to satisfy (\ref{conver}) are
\begin{eqnarray}
E\left[ \sum\limits_{i=1}^{n}\sqrt{\frac{c_{i}}{\sum\limits_{j=1}^{n}c_{j}}}%
\sqrt{U_{im_{i}}}\right] -E\left[ \sum\limits_{i=1}^{n}\sqrt{\lambda _{i}}%
\sqrt{V_{i}}\right] &\rightarrow &0  \label{cs1} \\
E\left[ \frac{\left( \sqrt{U_{im_{i}}}-\left( m_{i}+r_{i}\right) \sqrt{c_{i}}%
\right) ^{2}}{\left( m_{i}+r_{i}\right) \sqrt{\sum\limits_{j=1}^{n}c_{j}}}%
\right] &\rightarrow &0,~\forall i  \label{cs2}
\end{eqnarray}%
as $c\rightarrow 0$.

\begin{thm}
\label{th 241}Let $P$ a random policy satisfying $m_{i}\rightarrow +\infty
,~a.s.$, and suppose that condition (\ref{cond}) is true, then%
\[
E\left[ \sum\limits_{i=1}^{n}\sqrt{\frac{c_{i}}{\sum\limits_{j=1}^{n}c_{j}}}%
\sqrt{U_{im_{i}}}\right] -E\left[ \sum\limits_{i=1}^{n}\sqrt{\lambda _{i}}%
\sqrt{V_{i}}\right] \rightarrow 0, \ as\ c\rightarrow 0.
\]%

\end{thm}

\begin{pf}
Remark that
\begin{equation}
\lim_{m_{1},...,m_{n}\rightarrow +\infty }\sqrt{U_{im_{i}}}=\sqrt{V_{i}}%
,~a.s.  \label{conv}
\end{equation}%
Now%
\begin{eqnarray*}
\sup_{m_{1},...,m_{n}}E\left[ \left( \sqrt{U_{im_{i}}}\right) ^{2}\right]
&=&\sup_{m_{1},...,m_{n}}E\left[ U_{im_{i}}\right] \\
&=&E\left[ \psi ^{^{\prime \prime }}\left( \theta _{i}\right)
\prod\limits_{j\neq i}\psi ^{\prime ^{2}}\left( \theta _{j}\right) \right] \\
&=&E\left[ \psi ^{^{\prime \prime }}\left( \theta _{i}\right) \right]
\prod\limits_{j\neq i}E\left[ \psi ^{\prime ^{2}}\left( \theta _{j}\right) %
\right] <+\infty
\end{eqnarray*}%
hence, the uniform integrability of $\sqrt{U_{im_{i}}}$ follows from
condition (\ref{cond}) and martingales properties. Therefore, the
convergence in (\ref{conv}) holds in $L^{1}$ and consequently :%
\[
\sqrt{\frac{c_{i}}{\sum\limits_{j=1}^{n}c_{j}}}\sqrt{U_{im_{i}}}\rightarrow 
\sqrt{\lambda _{i}}\sqrt{V_{i}} ~ in ~ L^{1}, \ as\ c\rightarrow 0,
\]%
which achieves the proof.
\end{pf}

\section{The two stage procedure}

Following the previous section, our strategy now is to satisfy condition (%
\ref{cs2}). Then, we define the two stage
sequential scheme as follows.

\begin{description}
\item[Stage one] proceed for $k_{i}$ observation from population $P_{i}$ for 
$i=1,...,n$; such that $k_{i}\sqrt{c_{i}}\rightarrow 0$ and $%
k_{i}\rightarrow +\infty $ as $c_{i}\rightarrow 0$.

\item[Stage two] for $i=1,...,n$; select $m_{i}$ integer as follows :%
\[
m_{i}=\max \left\{ k_{i},\left[ \frac{\sqrt{U_{ik_{i}}}}{\sqrt{c_{i}}}-r_{i}\right]\right\}
\]%
where $\left[ x\right]$ denotes the integer part of $x$ and%
\[
U_{ik_{i}}=E\left[ \psi ^{^{\prime \prime }}\left( \theta _{i}\right)
\prod\limits_{j\neq i}\psi ^{\prime ^{2}}\left( \theta _{j}\right) /\mathcal{%
F}_{k_{1},...,k_{n}}\right]
\]
\end{description}
We give now the main result of the paper.
\begin{thm}
\label{th 251}Assume condition (\ref{cond}) satisfied for a $p\geq 1$, then the two
stage design is first order optimal.
\end{thm}

\begin{pf}
The $m_{i}$, as defined by the two stage, satisfies%
\[
\lim_{c_{i}\rightarrow 0}\left( m_{i}+r_{i}\right) \sqrt{c_{i}}=\sqrt{V_{i}}
\]%
and since 
\[
\sqrt{\sum\limits_{j=1}^{n}c_{j}}\left( m_{i}+r_{i}\right) =\frac{\sqrt{c_{i}%
}\left( m_{i}+r_{i}\right) }{\sqrt{\frac{c_{i}}{\sum\limits_{j=1}^{n}c_{j}}}}%
\rightarrow \sqrt{\frac{V_{i}}{\lambda _{i}}}, \ as\ c\rightarrow 0,
\]%
then%
\begin{equation}
\frac{\left( \sqrt{U_{im_{i}}}-\left( m_{i}+r_{i}\right) \sqrt{c_{i}}\right)
^{2}}{\sqrt{\sum\limits_{j=1}^{n}c_{j}}\left( m_{i}+r_{i}\right) }%
\rightarrow 0,~a.s.,\ as\ c\rightarrow 0  \label{ae}
\end{equation}%
To show the convergence in $L^{1}$, it will be sufficient to show the
uniform integrability of the left hand side of (\ref{ae}). So, observe that%
\begin{eqnarray*}
\frac{\left( \sqrt{U_{im_{i}}}-\left( m_{i}+r_{i}\right) \sqrt{c_{i}}\right)
^{2}}{\sqrt{\sum\limits_{j=1}^{n}c_{j}}\left( m_{i}+r_{i}\right) } &\leq &%
\frac{U_{im_{i}}+\left( m_{i}+r_{i}\right) ^{2}c_{i}}{\sqrt{%
\sum\limits_{j=1}^{n}c_{j}}\left( m_{i}+r_{i}\right) } \\
&\leq &\frac{U_{im_{i}}}{\sqrt{U_{ik_{i}}}}+\sqrt{U_{ik_{i}}}
\end{eqnarray*}%
$\sqrt{U_{ik_{i}}}$ is uniformly integrable, as a result of
martingales L$^{p}$ convergence properties with $p=2$. Now, remark that%
\[
\frac{U_{im_{i}}}{\sqrt{U_{ik_{i}}}}\leq \max_{k^{\prime }}\sqrt{%
U_{ik^{\prime }}}
\]%
and for the remainder of the proof, we use Doob's inequality to show that
$E\left[ \max_{k^{\prime }}\sqrt{U_{ik^{\prime }}}\right] <+\infty$.
We have,%
\[
E\left[ \max_{k^{\prime }}\left( \sqrt{U_{ik^{\prime }}}\right) ^{2p}\right]
\leq \left( \frac{2p}{2p-1}\right) ^{2p}E\left[ \left( \sqrt{V_{i}}\right)
^{2p}\right] <+\infty
\]%
hence, since $p\geq 1,$ $\max_{k^{\prime }}\sqrt{U_{ik^{\prime }}}$ is
integrable and the proof follows.
\end{pf}

\section{Conclusion}

The proof of the first order asymptotic optimality for the two stage design
has been obtained mainly through an adequate scaling of the approximated
Bayes risk associated with squared error loss and cost, a lower bound for
the scaled Bayes risk, martingales properties and Doob's inequality.


\bibliographystyle{sjs}

\section*{References}
\begin{thebibliography}{99}


\bibitem[Cohen and Sackrowitz (1984)]{cohen}
Cohen, A., Sackrowitz, H. B., 1984. Bayes double sample estimation procedures. Ann. Statist. 12(3), 1035--1049.

\bibitem[Hwang and Karunamuni (2008)]{hwang} 
Hwang, L.-C., Karunamuni, R. J., 2008. Asymptotically pointwise optimal allocation rules in Bayes sequential estimation. Statist. Probab. Lett. 78, 2490--2495


\bibitem[Hwang (1999)]{leng} 
Hwang, L.-C., 1999. Two-stage approach to Bayes sequential estimation in the exponential distribution. Statist. Probab. Lett. 45, 277--282.

\bibitem[Page (1987)]{page}
Page, C., 1987. Sequential design for estimating the product of parameters. Sequential Anal. 6(4), 351--371.

\bibitem[Rekab and Tahir (2000)]{rekab}
Rekab, K., Tahir, M., 2000. A two-stage sequential allocation scheme for estimating the product of several means. Stochastic Anal. Appl. 18(2), 289--298.

\bibitem[Shapiro and Wardrop (1980)]{shapiro}
Shapiro, C. P., Wardrop, R. L., 1980. Bayesian estimation for one parameter exponential family. J. Amer. Statist. Assoc. 75, 984--988.

\bibitem[Terbeche and Broderick (2005)]{terbeche aas}
Terbeche, M., Broderick, O., 2005. Two stage design for estimation of mean difference in the exponential family. Adv. Appl. Stat. 5(3), 325--339.

\bibitem[Terbeche (2000)]{terbeche phd}
Terbeche, M., 2000. Sequential design for estimation. Thesis (Ph.D.) - Florida Institute of Technology, Ann Arbor, MI, 74 pp. ISBN: 978-0493-00045-9

\bibitem[Terbeche et al. (2005)]{terbeche sort}
Terbeche, M., Broderick, O., Barbour, A., 2005. On sequential and fixed designs for estimation with comparisons and applications. SORT. 29, 217--234.


\bibitem[Woodroofe and Hardwick (2005)]{woodroofe}
Woodroofe, M., Hardwick, J., 1990. Sequential Allocation for an Estimation Problem with Ethical Costs. Ann. Statist. 18(3), 1358--1377.

\end{thebibliography}

\end{document}